\documentclass[sn-mathphys-num]{sn-jnl}

\usepackage{graphicx}%
\usepackage{multirow}%
\usepackage{amsmath,amssymb,amsfonts}%
\usepackage{amsthm}%
\usepackage{mathrsfs}%
\usepackage[title]{appendix}%
\usepackage{xcolor}%
\usepackage{textcomp}%
\usepackage{manyfoot}%
\usepackage{booktabs}%
\usepackage{algorithm}%
\usepackage{algorithmicx}%
\usepackage{algpseudocode}%
\usepackage{listings}%
\usepackage{epsfig}

\theoremstyle{thmstyleone}%
\theoremstyle{thmstyletwo}%

\theoremstyle{thmstylethree}%

\graphicspath{{figures/}}

\begin{document}

\title[Identifying bimodality in data using a proximity-based null model with an application to classifying cell cycle phases using oxidative stress]{Identifying bimodality in data using a proximity-based null model with an application to classifying cell cycle phases using oxidative stress}


\author[1]{\fnm{Michelle L.} \sur{Kovarik}}

\author[1]{\fnm{Tyler J.} \sur{Allcroft}}

\author*[2]{\fnm{Per Sebastian} \sur{Skardal}}\email{persebastian.skardal@trincoll.edu}

\affil[1]{\orgdiv{Department of Chemistry}, \orgname{Trinity College}, \orgaddress{\street{300 Summit St.}, \city{Hartford}, \postcode{06106}, \state{CT}, \country{USA}}}

\affil*[2]{\orgdiv{Department of Mathematics}, \orgname{Trinity College}, \orgaddress{\street{300 Summit St.}, \city{Hartford}, \postcode{06106}, \state{CT}, \country{USA}}}

\abstract{Detecting communities in large complex networks has found a wide range of applications in physical, biological, and social sciences by identifying mesoscopic groups based on the links between individual units. Moreover, community detection approaches have been generalized to various data analysis tasks by constructing networks whose links depend on individual units' measurements. However, identifying well-separated subpopulations in data sets, e.g., multimodality, still presents challenges as both community detection with existing null models and other partition methods either fail to give partitions that correspond to dips in the data or give partitions that do not correspond to dips in the data. Here we introduce a new spatially informed null model for this task that takes into account spatial structure but does not explicitly depend on distances between measurements. We find that community detection using this null model successfully identifies subpopulations in multimodal data and accurately does not for unimodal data. This new method represents a complement to statistical methods, as we treat data directly using a network science approach. We apply this new null model to first distinguish interphase and mitotic cell cycle phases and then S and G2 cell cycle phases in a group of \emph{Dictyostelium discoideum} cells using measurements of oxidative stress, which have been shown to correlate strongly with cell-cycle behaviors.}




\maketitle

\section{Introduction}\label{sec:01}

Community detection in complex networks has found applications in a wide range of applications by identifying functional and mesoscopic groups based on patterns of interactions and relationships between individual units~\cite{Girvan2002PNAS,Danon2005JStat,Newman2006PNAS,Fortunato2010PhysRep}. Examples include groupings in online social networks~\cite{Red2011SIAMRev}, partnerships in political institutions~\cite{Porter2005PNAS,Porter2007PhysA,Zhang2008PhysA,Macon2012PhysA}, architectures of neuronal organization~\cite{Bassett2011PNAS,Wymbs2012Neuron,Yin2020PNAS}, identifying antibiotic resistance in patients~\cite{Parker2015ACS}, and analysis of contact networks for sexually-transmitted diseases~\cite{Billock2020Sex}. In principle, these methods for identifying groups can be generalized to quantitative data sets by constructing networks from the data where nodes correspond to measurements and links are placed based on the pair-wise distances between measurements. These generalized methods for network-based data analysis, however, have their drawbacks, as community structures generically emerge in spatially embedded networks even when the data shows no signs of having significant subpopulations, i.e., multimodality or simply making erroneous partitions that do not correspond with any significant divide in the data~\cite{Gilarranz2020SciRep}. Thus, developing community detection frameworks that correspond more precisely with the identification of subpopulations, i.e., multimodality, and likewise accurately failing to do so in the absence of subpopulations, is an important task, as testing for multimodality in real datasets remains a challenging problem with few robust methodologies~\cite{Hartigan1985AnnStats}.

One such problems is the classification of cell cycle phases using measurements of oxidative stress. Reactive oxygen species are hypothesized to play important roles in microbial development of cells~\cite{Aguirre2005,Liu2020}. Reactive oxygen species are also linked with cell cycles, with mitotic cells having been shown to exhibit higher levels of oxidative stress than cells in interphase. Thus, subpopulations of cells distinguished by oxidative stress may correspond to different cell cycles. Given the practical challenges of taking individual cell measurements, linking cell cycle phases with measurements of oxidative stress could be useful in synthetic engineering tasks such as synchronizing cell cycles as well as developing a better understanding of the sources of heterogeneity between individual cells and throughout entire groups~\cite{Allcroft2023}. Here we seek to identify significant subpopulations in a group of \emph{Dictyostelium discoideum} cells using oxidative stress measurements quantified by the log-ratio of dihydrodichlorofluorescein diacetate (DCF) to carboxyfluorescein diacetate (CF). Technical details for the experimental measurments and data collection are provided in the Appendix.

To identify subpopulations in this and other datasets we utilize the modularity method for detecting communities in networks~\cite{Newman2004PRE}. This method is based on comparing an actual network structure to an appropriate null model that quantifies the expectation of the existence and strength of each pairwise link. Rather than using the classical Newman-Girvan null model~\cite{Newman2004PRE} or other null models that incorporates spatial structure by explicitly using distances~\cite{Expert2011PNAS,Simini2012Nature,Sarzynska2016CompNet,Cerina2012PLOSOne}, we introduce a new null model that is spatially informed but does not explicitly use distances. Thus, similar to the Newman-Girvan null model, this new null model uses primarily the local connectivity properties of nodes, but like distance-dependent null models a strong spatial structure is assumed without strong mixing. In essence, the null model is designed to assume that a node is most likely connected to the number of nodes matching its degree that are all close by, but this likelihood is not explicitly a function of any distance. We test this new null model on a series of synthetic data benchmarks to examine the effects of separation and asymmetry in the data, finding that it successfully identifies subpopulations, i.e., multimodality, when they are present and accurately fails to do so when they are not, i.e., the data is unimodal. We then turn to the task of partitioning cells according to their cell cycle phases using measurements of oxidative stress. Throughout our results, we will present the results from Hartigan's dip test~\cite{Hartigan1985AnnStats} as a point of comparison, but emphasize here that our new method should be viewed as a complement to this statistical test, as we aim to treat and make hypotheses about data directly.

The remainder of this paper is organized as follows. In Sec.~\ref{sec:02} we describe network-based approaches to identifying subpopulations in data using modularity and discuss the choice of null models. We then introduce a new spatially informed null model designed for identifying multimodality in data. In Sec.~\ref{sec:04} we evaluate this method with our new null model over a collection of synthetic data benchmarks. In Sec.~\ref{sec:05} we apply our method to real data taken from an ensemble of \emph{Dictyostelium discoideum} cells. Specifically, we use oxidative stress measurements first to identify cells in interphase vs mitosis, then in S and G2 phases. In Sec.~\ref{sec:06} we conclude with a discussion of our results.

\section{Data analysis, community detection, and the failure of existing null models}\label{sec:02}

Motivated by the task of identifying cell-cycle phases in experimental data of oxidative stress in \emph{Dictyostelium discoideum} cells, we consider data sets $\{x_i\}_{i=1}^N$ consisting of $N$ measurements with $i=1,\dots,N$. In our motivating example, each $x_i$ represents a relative measurement of oxidative stress in a cell, which is strongly correlated with the phase (interphase or mitosis) of the cell cycle. (More details on the data are given in Sec.~\ref{sec:05} and in Appendix~\ref{app:A}) In general we assume that each data point is a scalar, i.e., $x_i\in\mathbb{R}$, and therefore can be sorted in ascending order so that $x_1\le x_2\le\cdots\le x_N$. We then construct from this data set a network of $N$ nodes where each node $i$ corresponds to a measurement $x_i$ and links are placed between nodes based on the distance between corresponding measurements. Many choices can be reasonably made in terms of how links are placed, but here we consider the simple choice where a link is placed between nodes $i$ and $j$ if the measurements $x_i$ and $x_j$ satisfy $|x_i-x_j|\le r s_x$, where $r$ is a tunable local connectivity parameter and $s_x$ is the standard deviation of the dataset $\{x_i\}_{i=1}^N$. This network is described by the adjacency matrix $A$, whose entries $A_{ij} = 1$ if a link exists between nodes $i$ and $j$, and otherwise $A_{ij}=0$.

We then seek to map the task of identifying subpopulations in the data, i.e., multimodality, to finding communities in the corresponding network. Community detection has long been a topic of interest in the complex networks community and has a rich literature~\cite{Girvan2002PNAS,Danon2005JStat,Fortunato2010PhysRep}. Here we choose the modularity method~\cite{Newman2006PNAS}, where a given network structure is compared to an appropriately chosen null model and then partitioned into clusters that are more highly connected than expected by maximizing the so-called {\it modularity}. Mathematically, the modularity $Q$ can be expressed as
\begin{align}
Q=\frac{1}{2M}\sum_{i=1}^N\sum_{j=1}^N\left(A_{ij}-P_{ij}\right)\delta(c_i,c_j),\label{eq:01}
\end{align}
where $M$ is the total number of links in the network, $A$ is the adjacency matrix that describes the true network structure, the matrix $P$ represents the null model where the entry $P_{ij}$ gives the expected value of the link between nodes $i$ and $j$ according to the null model, $c_i$ and $c_j$ are the cluster (or community) membership of nodes $i$ and $j$, respectively, (typically communities are indexed $c_i=1,2,\dots,C$, where $C$ is the total number of communities) and $\delta$ is the Kronecker delta function so that $\delta(c_i,c_j)=1$ if nodes $i$ and $j$ are in the same community and $\delta(c_i,c_j)=0$ otherwise. Identifying communities in the network then corresponds to finding community memberships $c_i$ to sum over the most positive blocks in the modularity matrix $B=A-P$, thereby maximizing the modularity $Q$.

In general, finding communities to maximize modularity is a combinatorially difficult problem complicated by the fact that in typical cases the total number of communities in a network and number of nodes in each community are unknown. In our case this task is made significantly simpler given that we are first and foremost interested in finding a single partition of the network into two communities, corresponding to bimodality in the underlying data, and the ordering $x_1\le x_2\le\cdots\le x_N$ implies that a partition can be made using a single value $x_{\text{cut}}$ so that the two clusters correspond to values on either side of $x_{\text{cut}}$, i.e., $x_i\le x_{\text{cut}}$ implies that $i$ is in community 1 and $x_i> x_{\text{cut}}$ implies that $i$ is in community 2. Even for the task of identifying higher-order multimodality, i.e., identifying three or more communities, the ordering of the nodes allows this to be done relatively simply by identifying $C-1$ cut locations that separate $C$ contiguous groups.

Instead, our challenge here is in finding a null model (encapsulated in the matrix $P$) that allows us to find communities that accurately map to subpopulations, i.e., multimodality, in the original data. The classical choice for a null model was first presented by Newman and Girvan~\cite{Newman2004PRE} and is given by the expectation of a link being proportional to the product of the node-wise degrees, i.e., $P_{ij}=k_ik_j/(2M)$, where $k_i=\sum_{j=1}^N A_{ij}$ is the degree of node $i$. This null model is very effective when the network structure is well mixed and no other constraining information is known so that local connectivity i.e., node strength as measured by degree, are primarily important. In the case of spatially embedded networks, however, this choice generally yields spurious communities due to the lack of mixing and the strong spatial clustering of the network~\cite{Gilarranz2020SciRep}. A number of works have introduced and studied alternative null models designed specifically for spatially embedded network, for example the gravity model~\cite{Expert2011PNAS}, the radiation model~\cite{Simini2012Nature,Sarzynska2016CompNet}, and the exponential decay model~\cite{Cerina2012PLOSOne}, all of which similarly combine local nodal properties with information regarding the explicit distance between nodes. The gravity model, for instance~\cite{Expert2011PNAS}, uses a null model given by $P_{ij}=k_ik_j f(d_{ij})/Z$, where $d_{ij}=d(x_i,x_j)$ is the Euclidean distance between nodes $i$ and $j$, $f(d)$ is function that decreases monotonically as the distance $d$ increases, and $Z$ is a normalization constant set to ensure that $\sum_{i=1}^N\sum_{j=1}^NP_{ij}=\sum_{i=1}^N\sum_{j=1}^NA_{ij}$. In what follows we will use the gravity model with the choice $f(d_{ij})=\text{exp}(-d_{ij}^2)$, i.e., a Gaussian curve. Thus, these null models predict expected values of links in a way that balances the nodes' overall connectivity with their proximity in space.

\begin{figure}[t]
\centering
\epsfig{file =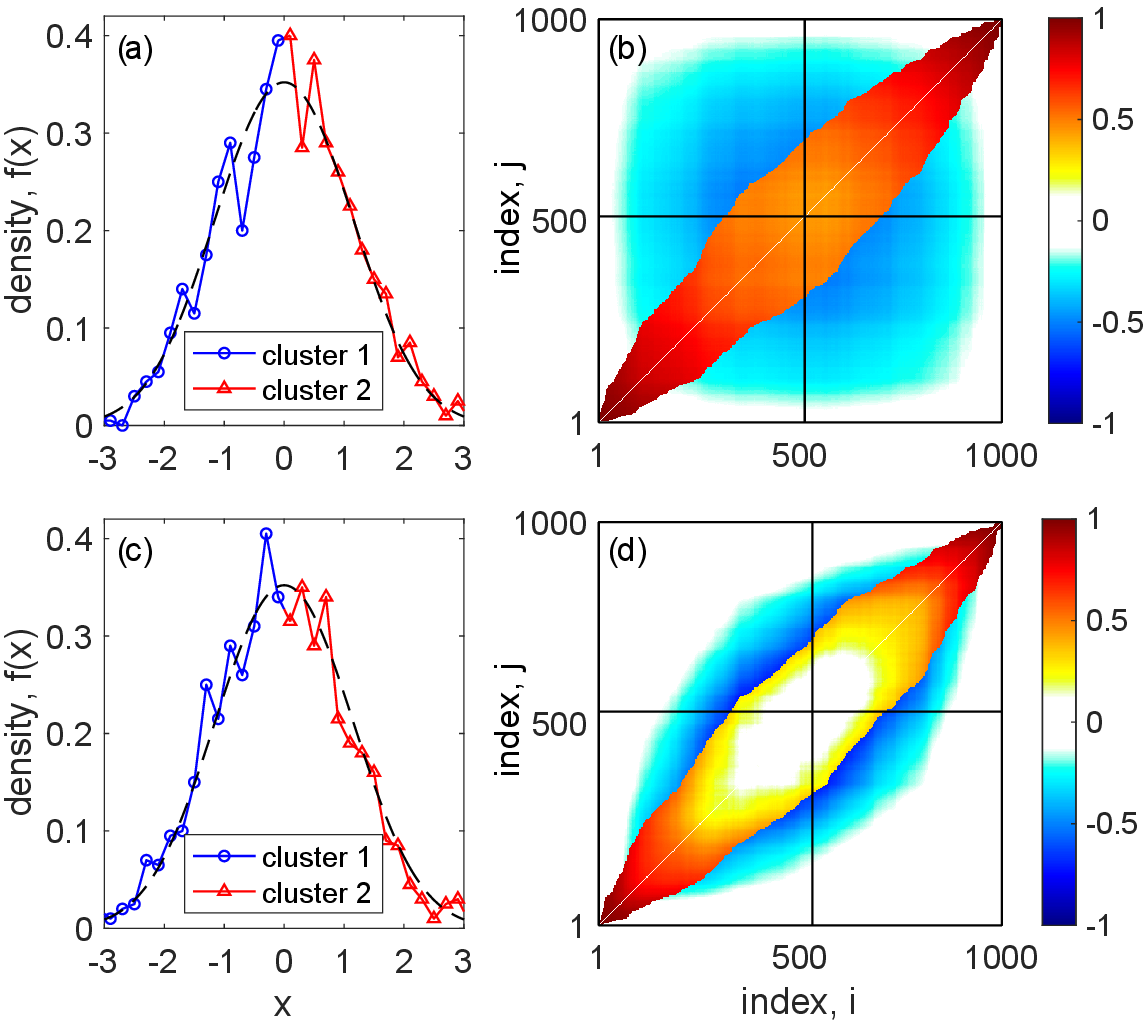, clip =,width=0.6\linewidth }
\caption{{\it Failure of existing null models to identify bimodality.} Two illustrative examples using a synthetic data set of size $N=10^3$ drawn from the density in Eq.~(\ref{eq:02}) with $\mu_1=-\mu_2=1/2$, $\sigma_1=\sigma_2=1$, and $\eta=1/2$ with network generated using $r=1/2$: (a) The distribution of distribution of data with clusters obtained using the NG null model indicated by blue circles and red triangles. The underlying distribution is plotted in dashed black. (b) A heat map of the corresponding modularity matrix with optimal partition indicated by the black lines. Analogous results using the gravity null model are given in panels (c) and (d). The optimal modularity in the two cases was found to be $Q=0.3705$ and $0.1814$, respectively.}\label{fig0}
\end{figure}

Using modularity to partition large ensembles of individual units into multiple coherent groups has been used in a wide range of applications~\cite{Red2011SIAMRev,Porter2005PNAS,Porter2007PhysA,Zhang2008PhysA,Macon2012PhysA,Bassett2011PNAS,Wymbs2012Neuron,Yin2020PNAS,Parker2015ACS,Billock2020Sex}. A wide range of other partitioning techniques exist as well, for instance, Laplacian eigenmaps~\cite{Belkin2003Neur}, diffusion maps~\cite{Coifman2005PNAS}, $k$-means~\cite{Lloyd1982}, Otsu's method~\cite{Otsu1979}, but in contrast to these methods the modularity method has the added component of assigning to any partition a measure of the quality of a given partition, i.e., the actual modularity value $Q$. However, for the task of identifying subpopulations that correspond to multimodality in data, existing methods fail. Specifically, community detection using existing null models, as well as the other methods referred to above, make strong partitions that do not correspond to bimodality. Take, for instance, a data set $\{x_i\}_{i=1}^N$ where each data point $x_i$ is identically and independently drawn from the following superposition of Normal distributions:
\begin{align}
g(x)=\frac{\eta}{\sqrt{2\pi\sigma_1^2}}e^{-\frac{(x-\mu_1)^2}{2\sigma_1^2}}+\frac{1-\eta}{\sqrt{2\pi\sigma_2^2}}e^{-\frac{(x-\mu_2)^2}{2\sigma_2^2}}.\label{eq:02}
\end{align}
Note that for the case of evenly-split clusters, $\eta=1/2$, equal variances, $\sigma_1^2=\sigma_2^2=\sigma^2$, and opposite means $\mu_1=-\mu_2=\mu$, this underlying distribution is bimodal if and only if $\mu>\sigma$. Thus, for the case of $\eta=1/2$, $\mu=1/2$, and $\sigma=1$ the underlying distribution is clearly unimodal. However, as an example of the failure of existing methods for this task and motivation for a new null model, we consider a data set of size $N=1000$ drawn from this choice of distribution and examine the optimal community partitions using both the classical NG null model~\cite{Newman2004PRE} as well as the gravity null model~\cite{Cerina2012PLOSOne}, where the underlying network is generated using $r=1/2$. In Fig.~\ref{fig0} we plot in panel (a) the distribution of the drawn data, with the two clusters identified using the NG null model in blue circles and red triangles (with the underlying distribution plotted in dashed black) and in panel (b) a heat map of the corresponding modularity matrix with the cut location indicated by black lines. In panels (c) and (d) we plot similar results for the gravity model, using $P_{ij}\propto k_ik_j\text{exp}(-|x_i-x_j|)$. Note that both methods make a strong partition near $x\approx0$ (the modularity for the NG and gravity cases were optimized at $Q=0.3705$ and $0.1814$, respectively), despite the clear lack of any bimodality in the data. These results are similar to those obtained by using Laplacian eigenmaps, diffusion maps, $k$-means (with $k=2$, and Otsu's method, all of which tend to favor a partition in the data near the middle, regardless of whether there is any bimodality in the underlying distribution. This failure of existing partitioning methods to faithfully correspond to multimodality thus motivates the need for a new null model that addresses this problem.

\begin{figure}[t]
\centering
\epsfig{file =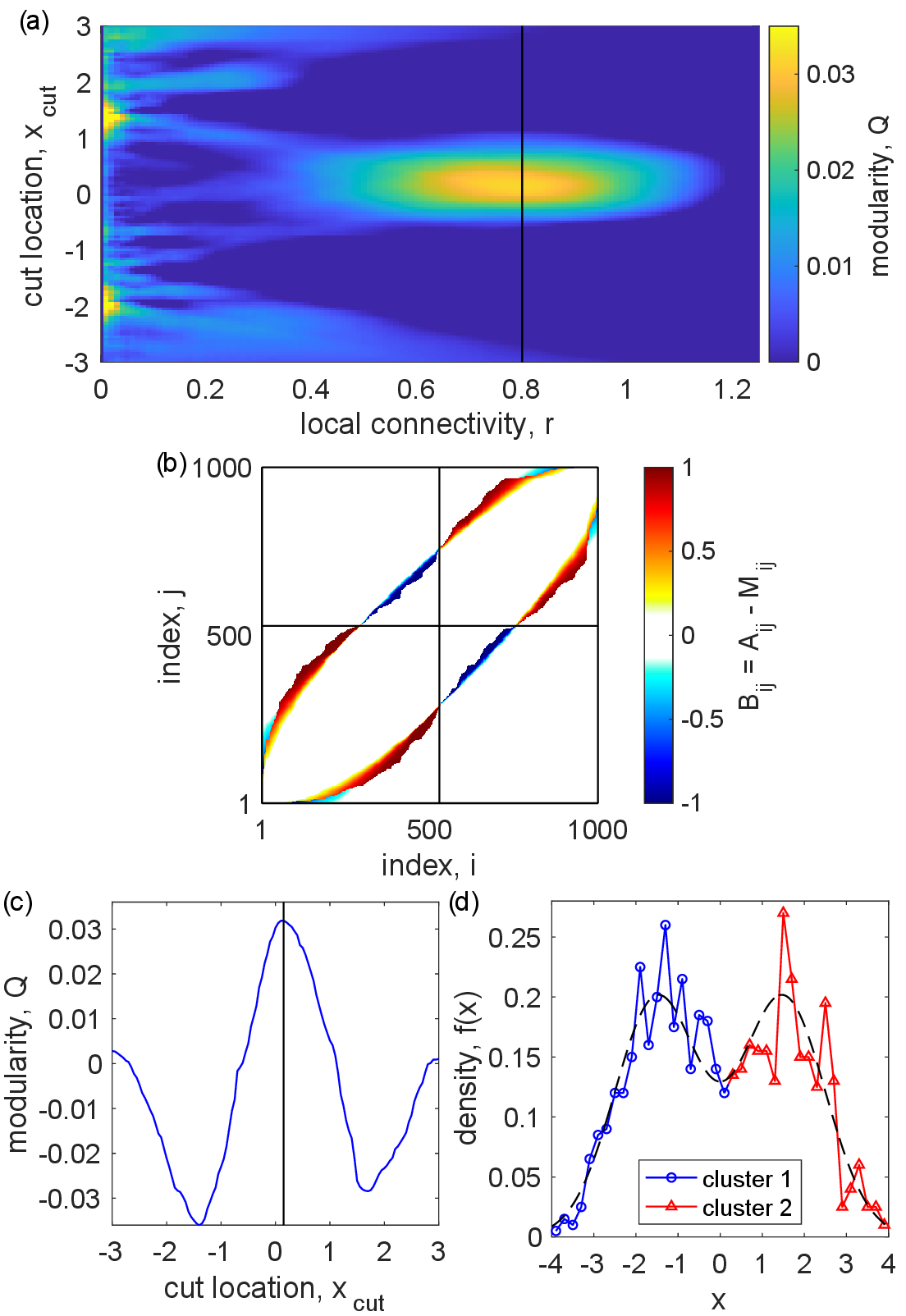, clip =,width=0.6\linewidth }
\caption{{\it Identifying subpopulations in data.} An illustrative example using a synthetic data set of size $N=10^3$ drawn from the density in Eq.~(\ref{eq:03}) with $\mu_1=-\mu_2=3/2$, $\sigma_1=\sigma_2=1$, and $\eta=1/2$: (a) A heat map of modularity as a function of the local connectivity parameter $r$ and the cut location $x_{\text{cut}}$. (b) A heat map of the entries of the modularity matrix using $r=0.8$ with vertical and horizontal lines indicated the cut index. (c) Also using $r=0.8$, the modularity $Q$ as a function of the cut location $x_{\text{cut}}$. (d) Lastly, the density of the raw data set with two main clusters indicated by blue circles and red triangles, as well as the underlying distribution plotted as a dashed black curve.}\label{fig1}
\end{figure}

\section{A new proximity-based null model}\label{sec:03}

In order to address the problem described above, here we introduce a new alternative via a null model that, as we will demonstrate, generates partitions that tend to correspond to bimodality in data far more faithfully than existing choices. The key to this new null model is, while it is spatially informed, it is not explicitly a function of the distances between nodes. To give some insight into its compositions, we note that for a network whose nodes are embedded in $\mathbb{R}$, we expect that a node $i$ with degree $k_i$ is most likely connected to the $k_i/2$ immediately to the left and the $k_i/2$ nodes immediately to the right. I.e., without knowing the details of the data set itself, there is no reason to expect that this symmetry is broken one way or the other. Then, to ensure that our new null model corresponds to an undirected network, we posit a link exists between nodes $i$ and $j$ with high likelihood if $|i-j|<\overline{k}_{ij}/2$ and low likelihood if $|i-j|>\overline{k}_{ij}/2$, where $\overline{k}_{ij}=(k_i+k_j)/2$. (Recall that the data is ordered so that indices satisfy $i\le j$ if $x_i\le x_j$.) We implement this simple idea using a sigmoidal function in the definition of the null model matrix to generate a transition between where we do and do not expect links to exist . Specifically, we set
\begin{align}
P_{ij} \propto \left[1+\text{exp}\left(\frac{|i-j|-\overline{k}_{ij}/2}{\frac{\alpha}{s_k}|k_i-k_j|}\right)\right]^{-1},\label{eq:03}
\end{align}
where $s_k$ is the standard deviation of the nodal degrees $\{k_i\}_{i=1}^N$ and $\alpha$ is a parameter that modifies the sharpness of the transition. Note that this sigmoidal function has a transition at $|i-j|=\overline{k}_{ij}/2$, so if the difference between the indices $i$ and $j$ is larger than half the mean of their degrees, i.e., there are many data points in between $x_i$ and $x_j$, then the likelihood of a link existing between these nodes is evaluated to be quite small. This transition becomes sharper for smaller $\alpha$. In the remainder of this work we consider $\alpha=10$ which tends to be relatively small compared to $s_k$ in the examples that follow, yielding a smooth but relatively sharp transition in the sigmoidal function. We note, however, that a relatively large range of choices for $\alpha$ offer similar results.

\begin{figure*}[t]
\centering
\epsfig{file =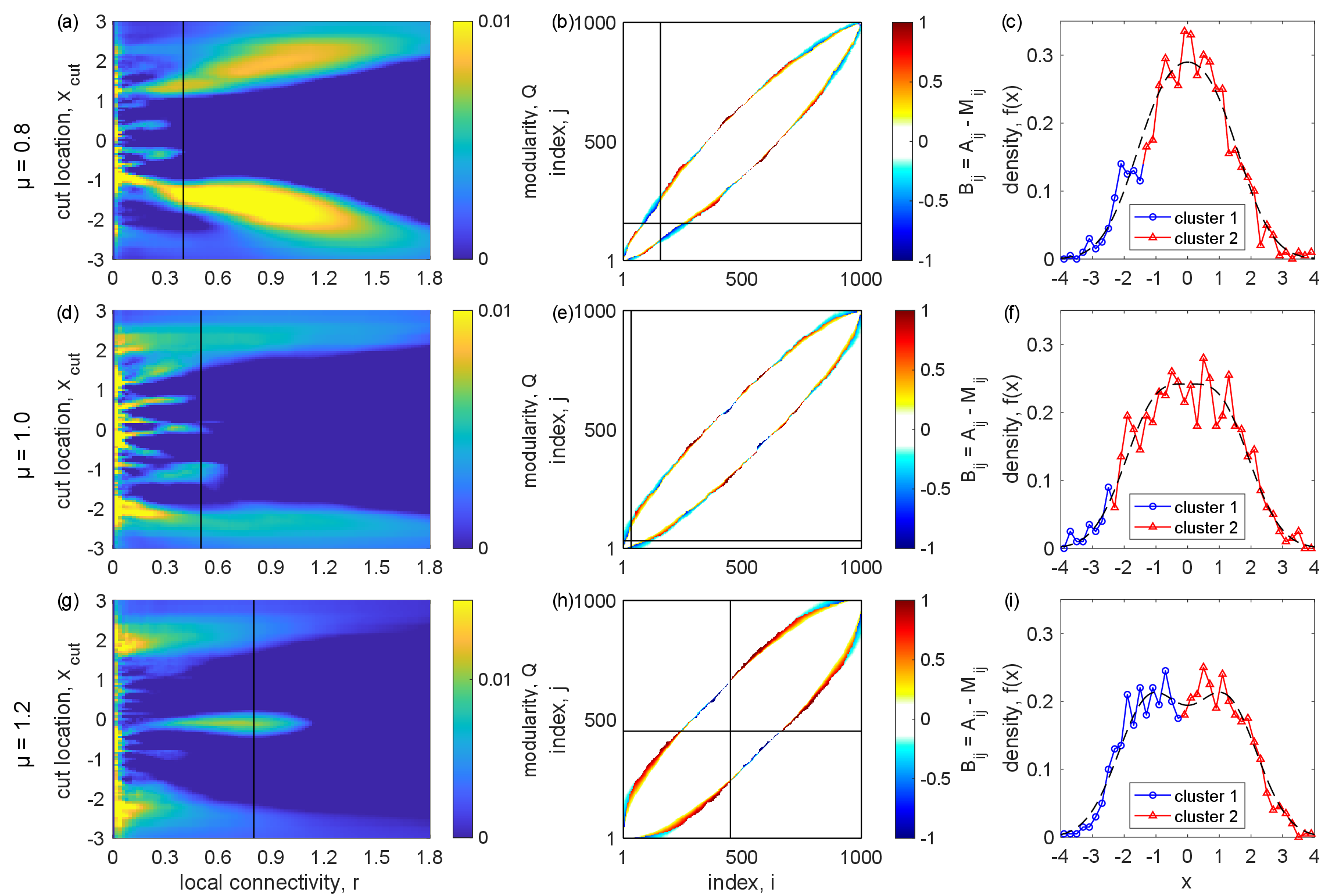, clip =,width=1.0\linewidth }
\caption{{\it Synthetic benchmarks: Effect of separation (part I).} For synthetic data set of size $N=10^3$ drawn from the density in Eq.~(\ref{eq:03}) with $\sigma_1=\sigma_2=1$, $\eta=1/2$ and varying $-\mu_1=\mu_2=\mu$, in each row we plot from left to right a heat map for $Q$ as a function of $r$ and $x_{\text{cut}}$, a heat map for the entries of the modularity matrix $B=A-P$, and the density of the raw data set with two clusters indicated by blue circles and red triangles, as well as the underlying distribution plotted as a dashed black curve. Here we plot the results for (a)--(c) $\mu=0.8$, (d)--(f) $\mu=1.0$, and (g)--(i) $\mu=1.2$.}\label{fig2}
\end{figure*}

To illustrate the use of this new null model we consider an illustrative example of a synthetic data set drawing from a the distribution $g$ given in Eq.~(\ref{eq:02}). Specifically, taking $\mu_1=-\mu_2=3/2$, $\sigma_1=\sigma_2=1$, and $\eta=1/2$ to ensure a moderate degree of bimodality, we draw $N=1000$ data points, construct networks over a range of the local connectivity parameter $r$, and for each $r$ calculate the modularity $Q$ over a range of the cut location $x_{\text{cut}}$. In Fig.~\ref{fig1}(a) we plot a heat map of $Q$ as a function of $r$ and $x_{\text{cut}}$, indicating larger values of $Q$ with yellow and smaller values with blue. Importantly, for a relatively wide range of $r$ values we find a significant local maximum of $Q$ at approximately $x_{\text{cut}}\approx0$, corresponding to a dip in the underlying distribution and the presence of bimodality. For the value $r=0.8$, as indicated by the vertical line in panel (a), we also plot the heat map of the modularity matrix in panel (b), indicating positive values with warm colors and negative values with cold colors. Here we can compare the structures of $A$ and $P$ directly, with positive values (red) corresponding to entries that exist in $A$ that are not expected by $P$ and negative values (blue) corresponding to values that are expected by $P$ but are absent in $A$. On the other hand, values near zero (white) correspond to locations where entries of $A$ are well-predicted by $P$, whether those links exist or not. In particular, the bimodality of the data results in a network with, as compared to the null model, more entries present towards the edges (lower left and upper right) and fewer present near the middle. In panel (c) we plot the modularity $Q$ vs the cut location $x_{\text{cut}}$ (also at $r=0.8$) where we see a clear maximum near $x_{\text{cut}}\approx0$ as indicated by the vertical black line. Specifically, the location of this local max location corresponding to the split between the two subpopulations in the data. Back in panel (b) we indicate the cut index with a horizontal and a vertical line, noting that this ``frames out'' the negative values in the modularity matrix. Note that this framing did not occur nearly as sharply in the examples shown in Fig.~\ref{fig0} that used the NG and gravity models. Lastly, in panel (d) we plot the density of the raw data categorized into two clusters (blue circles and red triangles), as well as the underlying density taken from Eq.~(\ref{eq:03}) plotted as a dashed black curve. We note that, while the bimodality of the underlying distribution is clear, (i) in practice the underlying distribution is typically unknown and (ii) identifying bimodality (both in terms of determining if the data is bimodal and, if so, where the split is) in observations of histograms of raw data in this and other examples is difficult due to sample-to-sample fluctuations.

\begin{figure*}[t]
\centering
\epsfig{file =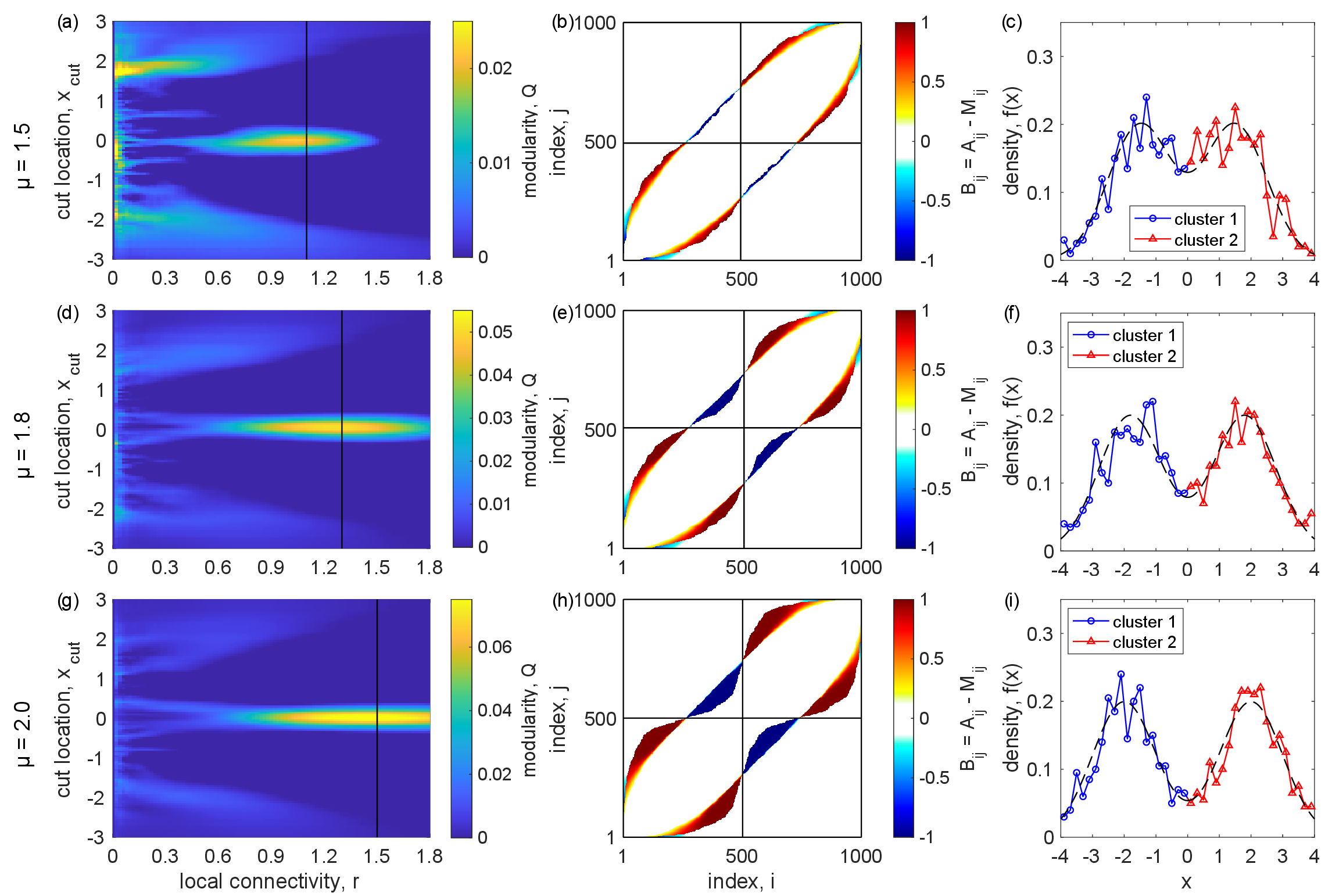, clip =,width=1.0\linewidth }
\caption{{\it Synthetic benchmarks: Effect of separation (part II).} Continued from Fig.~\ref{fig2}, the results for (a)--(c) $\mu=1.5$, (d)--(f) $\mu=1.8$, and (g)--(i) $\mu=2.0$.}\label{fig3}
\end{figure*}

This example highlights a few key steps in using the new null model for extracting subpopulations from data. First, the critical step in identifying subpopulations in a set of data comes in looking at a sweep of modularity over both the local connectivity parameter $r$ and the cut location $x_{\text{cut}}$. In particular, multiple subpopulations may only be identified if there is some robust range of $r$ for which one or more clear local maxima of the modularity persist. Moreover, by choosing a specific value of $r$ one might miss out on a significant regime. We note that this range does not have to occur at any specific $r$, but generically these values should not be too small. Second, at an appropriately chosen value of the local connectivity parameter, i.e., one in the range of $r$ where one or more clear maxima of the modularity exist, the data can be partitioned into subpopulations at a value of $x_{\text{cut}}$ that represents clear local maxima of $Q$. Similarly, the value of this local maximum may vary depending on the example, as long as it represents a significant local maximum as compared to the overall picture. Lastly, to visualize the clustering of data our null model it's also useful to examine the positive vs negative entries of the modularity matrix. Moving forward we test this process with our null model on a range of synthetic data benchmarks, then turn to our motivating application of identifying phases in cell cycles of \emph{Dictyostelium discoideum}.

\section{Synthetic data benchmarks}\label{sec:04}

To further evaluate the use of our spatially informed null model we examine a family of synthetic data benchmarks, testing performance over a range of data parameter values. Again we consider data sets drawn from the superposition of normal distributions given in Eq.~(\ref{eq:03}). We begin by testing the effect of separation between the two parts of the distribution. We note that for the symmetric case of $-\mu_1=\mu_2=\mu$, $\sigma_1=\sigma_2=\sigma$, and $\eta=1/2$, the distribution in Eq.~(\ref{eq:03}) is unimodal if $\mu\le\sigma$ and bimodal if $\mu>\sigma$. Thus, setting $\sigma=1$, the value $\mu=1$ is something of a critical value for performance testing: for $\mu\le1$ the use of our null model should not identify multiple subpopulations since the underlying data comes from a unimodal distribution, while for $\mu>1$ two subpopulations should be identified, corresponding to bimodality of the distribution, roughly separated at $x_{\text{cut}}=0$.

\begin{figure*}[t]
\centering
\epsfig{file =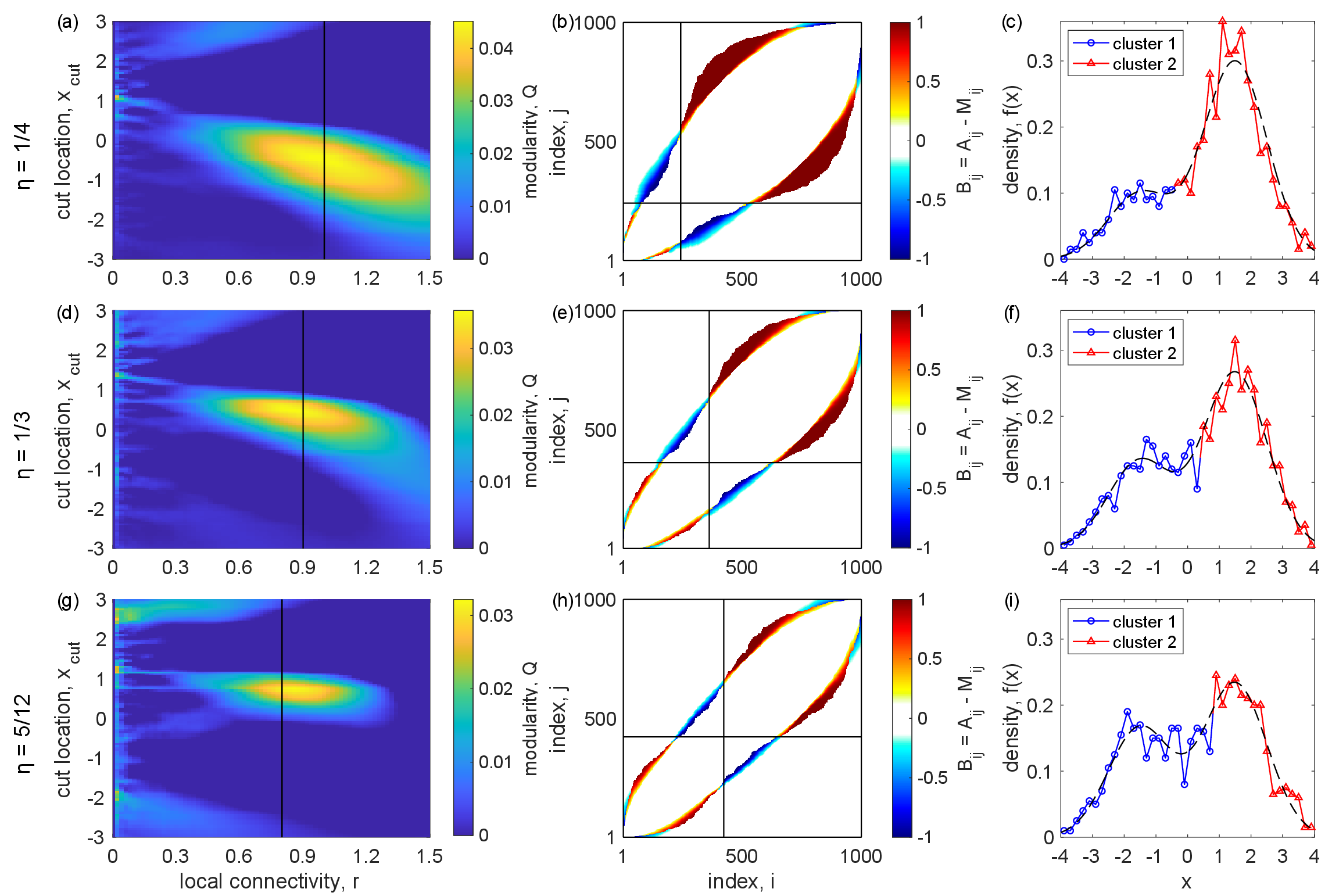, clip =,width=1.0\linewidth }
\caption{{\it Synthetic benchmarks: Effect of asymmetry.} For synthetic data set of size $N=10^3$ drawn from the density in Eq.~(\ref{eq:03}) with $-\mu_1=\mu_2=1.5$, $\sigma_1=\sigma_2=1$, and varying $\eta$, in each row we plot from left to right a heat map for $Q$ as a function of $r$ and $x_{\text{cut}}$, a heat map for the entries of the modularity matrix $B=A-P$, and the density of the raw data set with two clusters indicated by blue circles and red triangles, as well as the underlying distribution plotted as a dashed black curve. Here we plot the results for (a)--(c) $\eta=1/4$, (d)--(f) $\eta = 1/3$, and (g)--(i) $\eta=5/12$.}\label{fig4}
\end{figure*}

We consider six different cases corresponding to increased separation, i.e., stronger bimodality: $-\mu_1=\mu_2=\mu=0.8$, $1.0$, $1.2$, $1.5$, $1.8$, and $2.0$, in each case drawing a dataset of size $N=10^3$ with other parameters $\sigma_1=\sigma_2=\sigma=1$ and $\eta=1/2$. We proceed by sweeping over a range of the local connectivity parameter $r$, and cut location $x_{\text{cut}}$ for a heat map of the modularity $Q$, making an appropriate choice (if possible) for $r$ for a heat map of the entries of the modularity matrix $B=A-P$, and plotting the density of the raw data partitioned as best as possible into two subpopulations (blue circles and red triangles) along with the underlying density (black dashed curve). In Fig.~\ref{fig2} we plot the results for $\mu=0.8$ in panels (a)--(c), for $\mu=1.0$ in panels (d)--(f), and for $\mu=1.2$ in panels (g)--(i). We continue in Fig.~\ref{fig3} for $\mu=1.5$ in panels (a)--(c), for $\mu=1.8$ in panels (d)--(f), and for $\mu=2.0$ in panels (g)--(i). Beginning at smaller values of $\mu$ in Fig.~\ref{fig2}, the modularity heat maps for $\mu=0.8$ and $1.0$ [Fig.~\ref{fig1}(a) and (d)] show no significant local maxima of $Q$ near the bulk of the data. Instead, the most significant local maxima lie towards the periphery of the data. Choosing $r=0.4$ and $0.5$ for illustration, we can see that no significant portions of the modularity matrix share signs, implying no strong partition of the data in terms of modularity. In both cases the strongest partition separates only a tail of the data. Note that in both these cases no strong partition should be expected since the underlying distribution is unimodal. This is in contrast with the methods shown in Fig.~\ref{fig0}, which identified subpopulations even in a distribution without a clear presence of multimodality.

Moving to the modularity heat map for $\mu=1.2$ in Fig.~\ref{fig2}(g), we now see the emergence of a robust range of $r$ where a local maxima in $Q$ exists, near $x_{\text{cut}}=0$ between roughly $r=0.4$ and $1.0$. Choosing $r=0.8$ we see in panel (h) the bands of the modularity matrix now organize into coherent sections of positive and negative values, resulting in a stronger partition of the data, as illustrated in the two clusters in panel (i). The partition of data and identification of two subpopulations becomes stronger as $\mu$ is increased, as illustrated in Fig.~\ref{fig3}, with more robust bands of $r$ for which the heat maps of $Q$ yields a significant local maximum [panels (a), (d) and (g)] stronger bands of positive and negative vales in the modularity matrix [panels (b), (e), and (h)], and strong, accurate partitions of the data [panels (c), (f), and (i)]. Overall, using the new spatially informed null model successfully identifies subpopulations when the underlying data is bimodal ($\mu=1.2$, $1.5$, $1.8$, and $2.0$) and accurately does not when the underlying data is unimodal ($\mu=0.8$ and $1.0$). In addition to agreeing with the unimodality or bimodality of the known underlying distribution for these synthetic benchmark examples, we also report for each case the likelihood $p_{\text{dip}}$ of rejecting a null hypothesis of the data being unimodal using Hartigan's dip test~\cite{Hartigan1985AnnStats}: for $\mu=0.8$, $1.0$, $1.2$, $1.5$, $1.8$, and $2.0$ we find, respectively $p_{\text{dip}}=0.9886$, $0.6166$, $3.6\times10^{-3}$, $<1.0\times10^{-3}$, $<1.0\times10^{-3}$, and $<1.0\times 10^{-3}$, which match up well with our results.

\begin{figure*}[t]
\centering
\epsfig{file =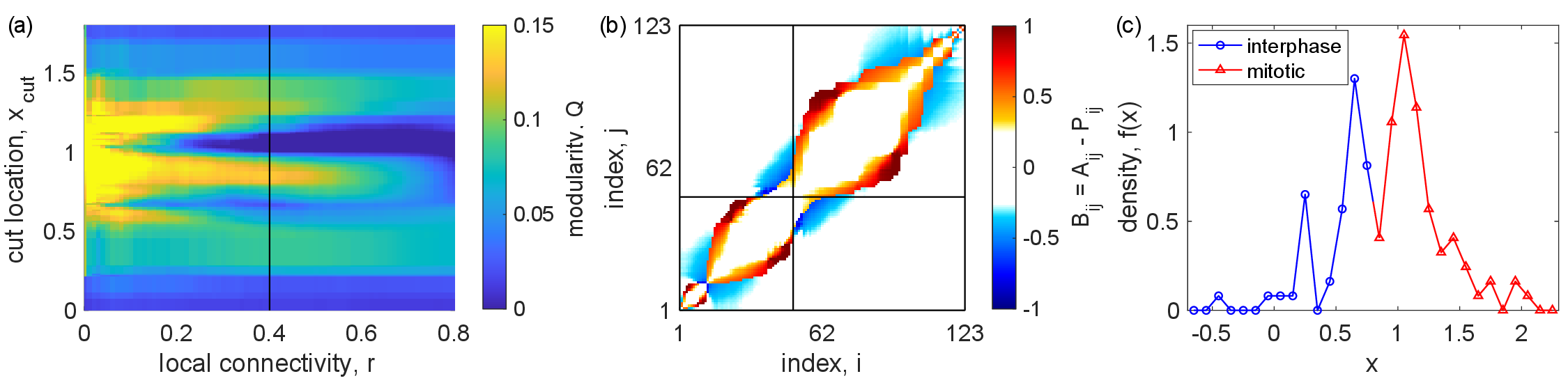, clip =,width=1.0\linewidth }
\caption{{\it Identifying interphase and mitotic subpopulations in a group of \emph{Dictyostelium discoideum} cells.} For measurements $x_i=\log(\text{DCF}_i/\text{CF}_i)$ from a group of $N=123$ \emph{Dictyostelium discoideum} cells 2 hours after release, (a) a heat map for $Q$ as a function of $r$ and $x_{\text{cut}}$, (b) a heat map for the entries of the modularity matrix $B=A-P$, and (c) the density of the raw data set with interphase and mitotic subpopulations indicated by blue circles and red triangles.}\label{fig5}
\end{figure*}

Before proceeding to real data we examine with another set of synthetic data the effect of asymmetry, i.e., subpopulations of different sizes. Specifically, taking the distribution in Eq.~(\ref{eq:03}) with $-\mu_1=\mu_2=1.5$ and $\sigma_1=\sigma_2=1$, we vary the asymmetry parameter $\eta$ and test our null model. In Fig.~\ref{fig4} we plot a heat map of $Q$ as a function of $r$ and $x_{\text{cut}}$, a heat map of the modularity matrix, and the histogram of the raw data partitioned into subpopulations (blue circles and red triangle) along with the underlying distribution (dashed black) for $\eta=1/4$ in panels (a)--(c), for $\eta=1/3$ in panels (d)--(f), and for $\eta=5/12$ in panels (g)--(i). (Note that the underlying distribution is bimodal in all three cases.) For all three examples the heat map of $Q$ reveals a robust range of $r$ with local maxima $Q$ [panels (a), (d) and (g)] and a strong partition of positive and negative bands exist in the modularity matrix taken at $r=1$, $0.9$, and $0.8$ [panels (b), (e), and (h)]. This leads to a successful identification of two subpopulation in all three examples of varying asymmetry [panels (c), (f), and (i)].

\section{Identifying cell states using oxidative stress}\label{sec:05}

Having tested our spatially informed null model on a range of synthetic data benchmarks, we now move to our motivating example of identifying populations of interphase vs mitotic \emph{Dictyostelium discoideum} cells using measurements of oxidative stress. Specifically, we consider a group of 123 cells which are prepared as follows. Cell cycles are first synchronized by chilling the cells overnight, then warming them, which induces all cells to proceed into mitosis. Upon release, two hours pass and measurements of dihydrodichlorofluorescein diacetate (DCF) and carboxyfluorescein diacetate (CF) are taken. At this two hour mark it is likely that a number of cells are in the mitosis phase of their cell cycle, but the relatively short duration of mitosis during the cell cycle (approximately 30 minutes out of an 8 hour cell cycle) and cell-to-cell heterogeneities throughout the population make it likely that some cells are mitotic while others are not. (Cells that are in interphase might have not yet reached mitosis or may have already gone through mitosis.) Measurements of DCF serve as overall measures of oxidative stress compared to an internal benchmark described by CF., so we consider for each cell $i$ the log ratio $x_i=\text{log}(DCF_i/CF_i)$ as a scalar-valued measure of oxidative stress.

\begin{figure}[t]
\centering
\epsfig{file =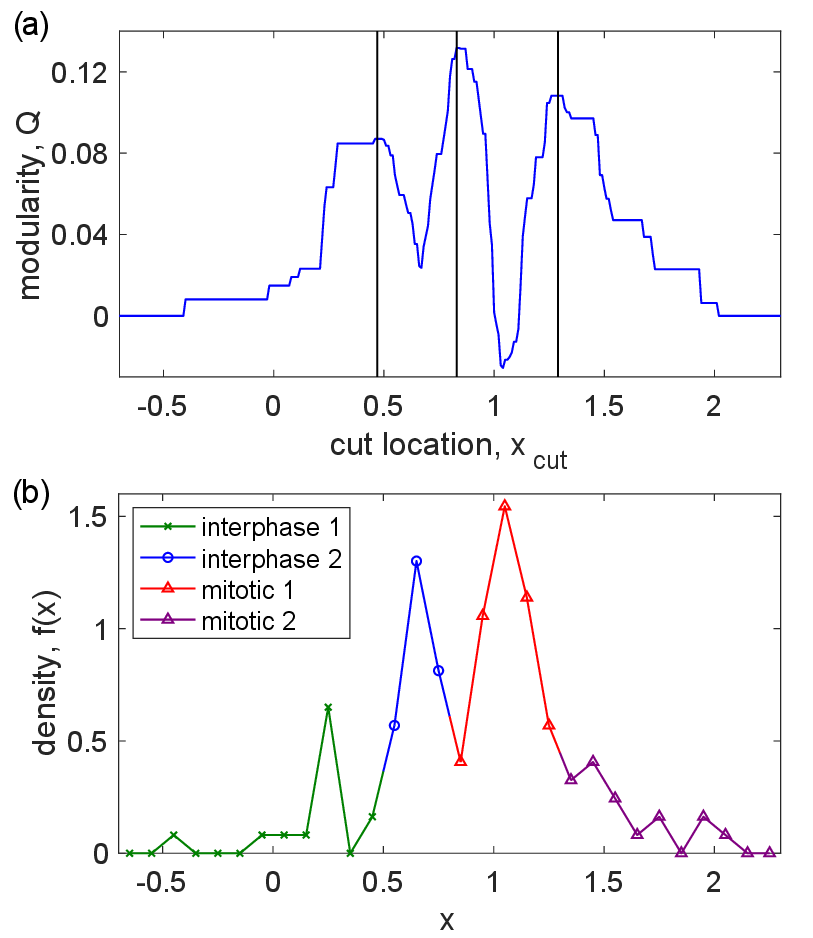, clip =,width=0.50\linewidth }
\caption{{\it Further partitions of subpopulations in a group of \emph{Dictyostelium discoideum} cells.} (a) For $r=0.4$ the modularity $Q$ as a function of cut location $x_{\text{cut}}$, with three local maxima indicated by vertical black lines. (b) Further partitioning of the data into four subpopulations using the three local maxima of $Q$ from panel (a).}\label{fig6}
\end{figure}

\begin{figure*}[t]
\centering
\epsfig{file =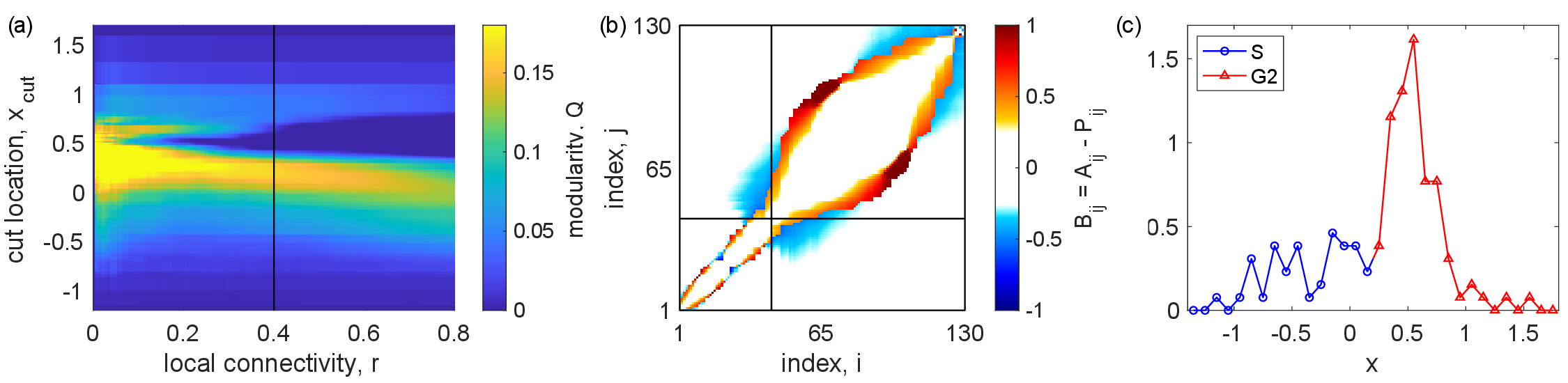, clip =,width=1.0\linewidth }
\caption{{\it Identifying interphase subpopulations in a group of \emph{Dictyostelium discoideum} cells: S and G2 phases.} For measurements $x_i=\log(\text{DCF}_i/\text{CF}_i)$ from a group of $N=130$ \emph{Dictyostelium discoideum} cells 5.5 hours after release, (a) a heat map for $Q$ as a function of $r$ and $x_{\text{cut}}$, (b) a heat map for the entries of the modularity matrix $B=A-P$, and (c) the density of the raw data set with S and G2 subpopulations indicated by blue circles and red triangles.}\label{fig7}
\end{figure*}

We proceed by building networks from this data set over a range of the local connectivity parameter $r$, and for each $r$ calculating the modularity $Q$ as the cut location $x_{\text{cut}}$ is varied. A heat map of $Q$ over $r$ and $x_{\text{cut}}$ is plotted in Fig.~\ref{fig5}(a). In fact, we see three somewhat coherent local maxima for $Q$ over ranges of $r$, but one that is the most prominent and stretches the furthest, up to approximately $r=0.6$. Taking $r=0.4$ we plot a heat map of the entries of the modularity matrix in panel (b), and the density of the raw data partitioned into inferred subpopulations representing interphase and mitotic cells in blue circle and red triangles in panel (c). The strength of these results in light of the synthetic data benchmarks presented above suggest that, given the strong correlation between oxidative stress and cell cycle phases, the upper and lower subpopulations identified are likely good partitions of the cells into interphase and mitotic groups.

We also look more closely at the other local maxima of $Q$ that emerge over ranges of $r$. In Fig.~\ref{fig6}(a) we plot $Q$ as a function of $x_{\text{cut}}$ at $r=0.4$, indicating the three local maxima with vertical black lines. The local maximum used above to separate interphase and mitotic subpopulations corresponds to that in the center, but the other two also correspond to nearly as large values of $Q$. making additional partitions at corresponding values effectively split the interphase and mitotic subpopulations into two subpopulations each, yielding in total four subpopulations (interphase 1, interphase 2, mitotic 1, and mitotic 2), as illustrated in the density of the data in panel (b). However, returning to the heat map of the modularity matrix in Fig.~\ref{fig5}, these additional splits correspond to the areas of negative entries near the periphery. It is difficult to say whether these constitute truly different subpopulations, especially since these portions of the modularity matrix appear to be somewhat similar to the small bands of negative values near the periphery in our synthetic data sets, e.g., see Fig.~\ref{fig3}(b). These additional splits may correlate to biological sub-phases of interphase and mitosis or may be the result of fluctuations and limited data.

Inspired by the possibility of identifying addition sub-phases of the data, we close by examining a new data set taken from the same group of cells, but 3.5 hours later on, i.e., with DFC and CF measurements taken at 5.5. hours after release. At this point it is likely that all cells are no longer mitotic and are in some sub-phase of interphase, specifically either the S or G2 phase. (Note that S and G2 correspond to the phases where, respectively, the cells are either replicating DNA and synthesizing protein in preparation for the next mitosis phase. Moreover, \emph{Dictyostelium discoideum} does not pass through a G1 phase.) Importantly, S phase is much shorter than G2 phase~\cite{Weeks1994}, and therefore is likely to correspond to a smaller group of cells at any given time. Furthermore, investigations into cancer cells show that DCF/CF ratios measuring oxidative stress tend to be smaller in S phase as compared to G2 phase~\cite{Patterson2019}. 

We repeat our approach from above on this new set of data and present the results in Fig.~\ref{fig7}. In Fig.~\ref{fig7}(a) we observe a significant local maximum for the modularity across a wide range of the local connectivity parameter $r$, indicating the possibility of multiple subpopulations. For $r=0.4$ we present the heat map of the modularity matrix in Fig.~\ref{fig7}(b). Using the optimal partition obtained from this value of $r$, denoted by the solid lines in Fig.~\ref{fig7}(b), we plot the density of the raw data in Fig.~\ref{fig7}(c). Note that modularity optimization used with our new null model identifies two subpopulations, one of which is smaller and has lower values of the DCF/CF ratio. Thus, it is plausible that the two subpopulation identified correspond to S and G2 phases of the post-mitotic cells, with S phase cells having divided later or more slowly than the G2 cells due to intrinsic or external heterogeneities. This provides a testable biological hypothesis, given a suitable experimental setup for monitoring and taking measurements of individual cells. It is worth noting that for both the 2 hour data (i.e.f, Figs.~\ref{fig5} and \ref{fig6}) and the 5.5 hour data (i.e., Fig.~\ref{fig7}) Hartigan's dip test is inconclusive, yielding p-values of approximately $0.23$ and $0.98$, suggesting that community detection with our new null model may extract structure that would otherwise not be detected.

\section{Discussion}\label{sec:06}

In this work we've considered the task of identifying subpopulations in data corresponding to multimodality and introduced a new null model for doing so via community detection using modularity maximization. This null model is proximity-based, as it assumes connectedness between adjacent data points, but, unlike previous spatially-dependent null models used for community detection, we do not explicitly use distances. We tested this new approach on a family of synthetic data benchmarks where the bimodality of a data set can be systematically tuned, finding success as both the separation of two subpopulations, as well as the asymmetry between the two subpopulations, were varied. Our motivating application is identifying subpopulations in groups of \emph{Dictyostelium discoideum} cells via measurements of oxidative stress. Specifically, we are able to separate groups of cells that plausibly fall into mitosis and interphase states, as well as cells that fall into S and G2 states. In the data set collected (consisting of $N$ measurements of oxidative stress from different cells) the approach produces compelling results and a testable biological hypothesis. 

As this work focuses on the simple case of univarate data, a key step for future work will be in extending this null model to multivariate data. While the construction of a network from a multivariate data set is not a difficult task, the generalization of the rule used in this paper for defining the null model [i.e., Eq.~(\ref{eq:02})] is not trivial. One possible choice would be to use an adaptive $k$-nearest neighbors rule for the null model, where the expected link between two nodes depends on the overlap between each node's $k_i$ nearest neighbors. Moreover, the task of parameterizing different network partitions in multiple dimensions will be more complicated and thus will necessarily draw on other community detection approaches for spatial networks. The extension of identifying subpopulations from one-dimensional data to multi-dimensional data could reveal important new approaches for identifying different phenotypical states that are difficult to observe using a single variable.

\bmhead{Author contributions}
MLK, TJA, and PSS performed all work associated with this manuscript.

\bmhead{Funding}
This work was funded by NSF grant MCB-2126177 (MLK and PSS).

\bmhead{Availability of Data and Materials}
Data and materials are available on reasonable request from PSS.

\section*{Declarations}
\subsection*{Competing Interests}
The authors declare that they have no conflict of interest.

\begin{appendices}

\section{Experimental measurements and data collection}\label{app:A}

The data set considered above in Sec.~\ref{sec:05} consists of \emph{Dictyostelium discoideum} cells for which we consider the measurements of oxidative stress. To quantify oxidative stress, we compare the levels of fluorescence from dihydrodichlorofluorescein diacetate (DCF) to carboxyfluorescein diacetate (CF) after initially loading each cell with $250$ $\mu$M of each. In particular, CF acts as an internal standard while DCF is a reactive oxygen species probe, indicating oxidative stress. Additionally, each cell is also initially loaded with 40 $\mu$M Rose Bengal as a photosensitizer to induce oxidative stress. DCF is a common indicator for reactive oxygen species~\cite{Chen2010}. Recent research suggests that ratioed measurements with CF more accurately reflect cell-to-cell variation in reactive oxygen species by accounting for spurious sources of heterogeneity~\cite{Metto2013,Rodogiannis2018,Sibbitts2020,Allcroft2023}

We note that heterogeneities in the individual measurements of CF and DCF are significant, so overall oxidative stress is best gauged using their log-ratio, yielding a single value for each cell.  After incubation and washing in low-fluorescence media, cell are exposed to a blue LED light source for a time ranging between 0-10 minutes, after which the log ratio $x=\text{log}(\text{DCF}/\text{CF})$ is measured as an individual measurement of oxidative stress. Additional details can be found in Ref.~\cite{Allcroft2023}.

\end{appendices}

\bibliography{bibliography}

\end{document}